\documentstyle[aps,epsf]{revtex}  
\global\firstfigfalse 
\global\firsttabfalse  
\begin{document}        
\baselineskip 14pt

\title{The Ratio of Dimensionless Jet Cross Sections at the Tevatron}

\author{J. Krane$^{1}$}
\address{$^{1}$Iowa State University, Ames, IA 50011}
\maketitle

\begin{abstract}
Several sources, both experimental and theoretical, subject inclusive jet 
cross section measurements to large uncertainties.  On the experimental side,
the energy scale contributes between 10\% and 30\% uncertainties.  On the 
theoretical side, choice of parton distribution function introduces a
20\% variation in the prediction; freedom in renormalization scale provides
another uncertainty of up to 30\%.  The ratio of inclusive jet cross
sections reduces the uncertainty from these major sources, permitting a
very precise test of next--to--leading order QCD.  The preliminary results
from D\O\ and CDF differ from simple QCD predictions in normalization
and differ from each other at small values of jet $x_T$.  
\end{abstract}

\section{Introduction}
At sufficiently high energy, field theory calculations reduce the
complexities of proton--antiproton interactions to simple scattering
processes involving only one constituent from each particle. Identifying
these constituents, called ``partons,'' as quarks and gluons, quantum
chromodynamics (QCD) calculations determine the matrix elements for a given
scattering process, but require input from empirically-determined parton
momentum distribution functions (pdf's). Perturbative QCD calculations
provide specific predictions for the rates of scattering processes as a
function of final--state parton momentum and angle, but the unknown
initial--state momentum superimposes a boost to the final
system.  Consequently, cross section predictions reflect an average over the
initial momenta through sampling the pdf over a large number of events.
The two largest uncertainties in the QCD prediction result from the choice
of pdf and the choice of renormalization scale. 

Experimentally, a final state parton is observed as a shower of collimated
particles: a jet. Jet cross sections are reported in terms of $E_{T}$, 
the jet
energy in the plane transverse to the beamline. This analysis compares the 
production rate of jets at two center--of--mass energies, 630 and 1800 GeV, 
for two reasons.  First, the ratio of cross sections
constrains experimental uncertainties that are common to 
both studies. Second, the ratio suppresses the QCD prediction's 
sensitivity to the choice of pdf. 
The ratio of cross sections thus provides a stronger test of the matrix 
element portion of the calculation than a single cross section 
measurement alone.

\section{Context}
A simple parton model without gluon bremsstrahlung would predict a jet cross
section that scales with center--of--mass energy ($\sqrt{s}$). In this
scenario, a dimensionless jet cross section, 
$E_{T}^{4}\cdot E\frac{d^{3}\sigma }{dp^{3}}$, plotted as a function of jet 
$x_{T}\equiv \frac{E_{T}}{2\sqrt{s}}$, would remain constant with respect to 
changes in center--of--mass energy.

Although past cross section ratio analyses 
at lower $\sqrt{s}$ exhibited significant deviation from the 
scaling model, the dimensionless framework continues to provide a useful
context for comparisons with the more developed QCD theory; thus, the prior
tests motivate the selection of variables in this work. The time--integrated
luminosity provides a measure of the statistical power of these earlier
cross section measurements: the 1985 results from UA2 \cite{ua2_ratio} 
include 122 nb$^{-1}$ ($\sqrt{s}=546$ GeV) and 310 nb$^{-1}$ 
($\sqrt{s}=630$ GeV) of data, 
whereas the 1993 result from CDF \cite{cdf_ratio} compared 8.58 nb$^{-1}$ 
($\sqrt{s}=546$ GeV) to 4430 nb$^{-1}$ ($\sqrt{s}=1800$ GeV).

As reported in \cite{xs_1800}, the D\O\ collaboration at Fermilab measured
the inclusive jet cross section at $\sqrt{s}=1800$ GeV  using 93,000 nb$^{-1}
$ of data. This article presents a complementary inclusive jet cross section
analysis at $\sqrt{s}=630\;\mathrm{GeV}$, with 537 nb$^{-1}$ of data.  The 
most recent analyses from the CDF collaboration 
at Fermilab use data samples of similar size at both center--of--mass energies.
Because in each case, the data sets for both center--of--mass energies were 
collected with the same detector \cite{detector}\cite{CDFdetector}, many 
uncertainties in the results are highly--correlated, particularly uncertainties
in the calorimeter response and calorimeter noise.  Additionally, the 
explicit expression of the correlations in uncertainties from bin to bin 
provides a more stringent comparison to QCD than was previously possible.

The remainder of this article focuses on the methods of the D\O\ analysis
of the jet cross section and ratio, reserving discussion of the CDF result
for last.  Although the details differ, both experiments require the same 
components for their respective analyses: data reconstruction, sample 
selection, energy scale correction, and resolution correction.

\section{Data Reconstruction and Sample Selection}
The D\O\ collaboration employs the variables transverse energy $\left(
E_{T}\right) $ and pseudorapidity $\left( \eta =-\ln \left( \tan \frac{%
\theta }{2}\right) \right) $ to describe the hadronic particle showers that
result from inelastic p\={p} collisions. Here, $\theta $ measures the polar
angle relative to the proton beam direction.
Jets are defined by their hadronic shower $E_{T}$ and $\eta -\phi $
centroid. Starting with the most energetic deposits in the calorimeter, the
D\O\ reconstruction algorithm \cite{reco} iteratively finds the
energy--weighted centroid for each hadronic shower within a cone of
dimensionless radius 0.7. When two such showers overlap, they are merged\
into a single jet if they share more than 50\% of the $E_{T}$ of the lowest--%
$E_{T}$ shower; otherwise, they are split into two separate jets, each with
its own centroid and $E_{T}$.

The online data selection procedure triggers on events that contain at
least one jet above threshold. A small correction, $3\%$ for the first data
point and decreasing thereafter, removes the effect of a slight trigger
inefficiency at low $E_{T}$. The offline data selection procedure, which
eliminates background caused by electrons, photons, noise, or cosmic rays,
closely follows the methods \cite{daniel}\cite{krane} of the 1800 GeV
analysis. The efficiency of jet selection is nearly constant as a
function of jet $E_{T}$, approximately $96\%$. To maintain optimal precision
in jet energy measurements, a vertex requirement discarded jets originating from
interaction points more than 50 cm from the longitudinal center of the
detector, lowering the total data selection efficiency to $82\%$. The cross
section uncertainty associated with all efficiencies amounts to less than $%
0.5\%$.

\section{Energy Scale and Resolution Corrections}
The jet energy scale correction, described in \cite{escale}, corrects the
energy response of the D\O\ calorimeter and removes cone boundary effects,
multiple interaction effects, calorimeter noise, and the underlying event
energy resulting from spectator partons. The response correction increases\
the $E_{T}$ of jets by $22\%$ when the raw calorimeter $E_{T}$ is $20\;%
\mathrm{GeV}$, and increases the $E_{T}$ by $10\%$ at asymptotically--high raw $%
E_{T}$'s. Calorimeter noise contributes on average $1.6\;\mathrm{GeV}$ of $%
E_{T}$ to each jet. \ The underlying event, which represents the only
significant difference between center--of--mass energies, contributes $0.6\;%
\mathrm{GeV}$ to each jet at $\sqrt{s}=630\;\mathrm{GeV}$, compared to $0.9\;%
\mathrm{GeV}$ at $\sqrt{s}=1800\;\mathrm{GeV}$. Simply speaking, a
jet's measured $E_{T}$ increases by 12 to 14\% after the total
energy scale correction. Uncertainties from the noise and response
corrections dominate the systematic uncertainty band of the final result.

Although the energy scale algorithm corrects jets from their reconstructed $%
E_{T}$ to their ``true'' $E_{T}$ on average, the energy scale cannot remove
the random fluctuations of individual jets about this average value. The
resulting imperfect resolution of a jet's $E_{T}$ about its true value leads
to a smearing effect that effectively inflates the observed cross section,
especially in the steepest portions of the distribution. After removing 
contributions that do not result from detector effects, deviations from
perfect momentum balance in dijet events provide a measure of the average
resolution as a function of jet $E_{T}$.

The observed cross section distribution can be described as the convolution
of the resolution function and an initial distribution, 
$F(E_{T})=\int G(E_{T}-E_{T}^{\prime })\cdot f(E_{T}^{\prime })\;
dE_{T}^{\prime }$, where $G$ is a Gaussian distribution and the initial 
distribution $f$ is modelled by: 

\begin{equation}
f(E_{T})=A(E_{T})^{B}(1-\frac{E_{T}}{\sqrt{s}})^{D}\text{.}  \label{eq_lilf}
\end{equation}

The Gaussian has $E_{T}$--dependent width, and represents the jet $E_T$ 
resolution. The three parameters of the ansatz function, $A$,
$B$, and $D$,
are determined with a best--fit of the smeared distribution, $F$, to the
observed cross section. The correction factor for the smearing effect is
expressed as $C(E_{T})=f(E_{T})/F(E_{T})$; the measured cross section is
multiplied by $C$ on a bin--by--bin basis. The uncertainty of unsmearing
the cross section is expressed in an error matrix, where the partial
derivatives of the correction factor with respect to each parameter are
determined numerically.  In addition to the fitting
uncertainties of $C$, this matrix includes contributions from the 
resolution determination.

\begin{figure}[f]      
\centerline{\epsfxsize 7.0 truein \epsfbox{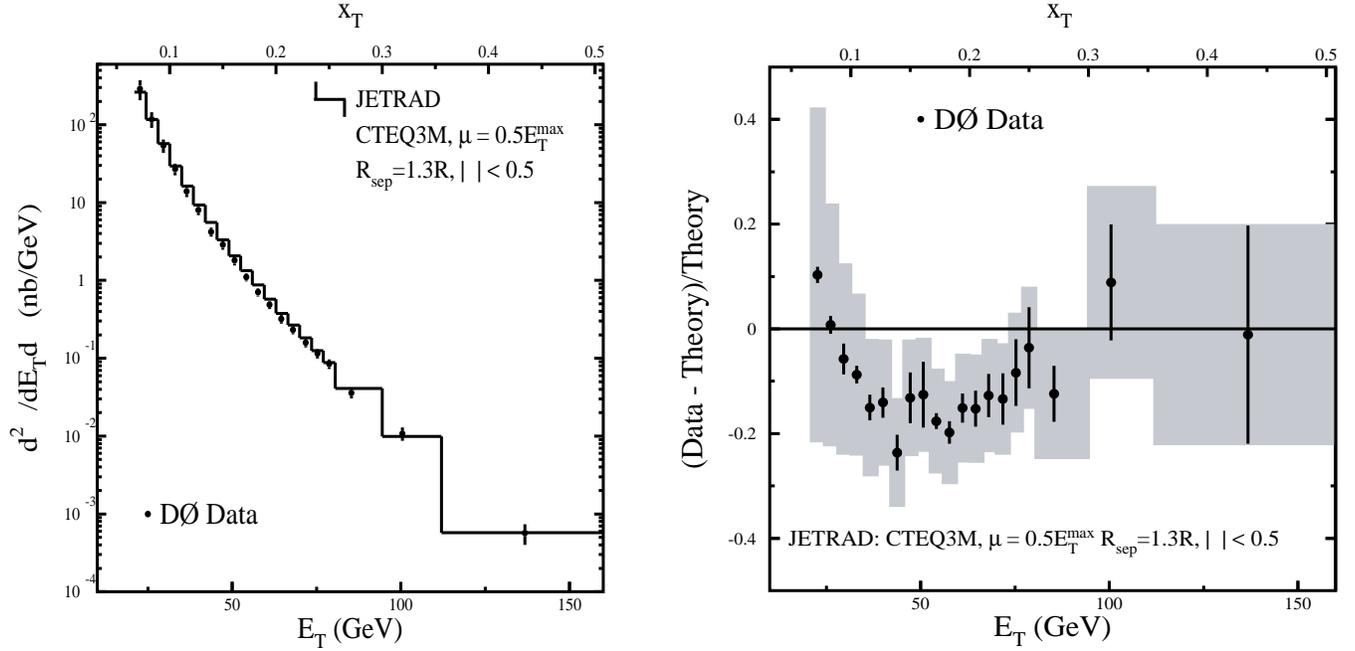}}   
\caption[]{(Left) The inclusive jet cross section at $\sqrt{s}=630$ GeV.
Black points indicate the cross section measurement and statistical errors;
histogram line indicates NLO QCD Prediction from \textsc{Jetrad}.
(Right) Fractional deviation of the cross section data and NLO QCD.  Shaded
blocks indicate systematic uncertainty.}
\label{fig:xsecs}
\end{figure}

\section{Results}
The unsmeared $\sqrt{s}=630\;\mathrm{GeV}$ jet cross section appears in
the left side of Figure \ref{fig:xsecs}. The point position within 
each $E_{T}$ bin occurs where
the ansatz takes its average value within the bin. The corresponding $x_{T}$
positions appear on the upper axis in anticipation of the ratio of cross
sections; by design, the $x_{T}$ points provide an exact match to the points
derived from the jet cross section at $\sqrt{s}=1800\;\mathrm{GeV}$.
The right side of Figure \ref{fig:xsecs} compares the cross 
section to a NLO QCD prediction,
with renormalization and factorization scales set to $E_{T}/2$ and using
CTEQ3M parton distribution function. $R_{sep}$, a phenomenological parameter 
\cite{Rsep} in the prediction, was set to the usual value of 1.3. The shaded
blocks in the figure indicate the $1\sigma $ systematic uncertainty of the
cross section measurement, while the vertical bars indicate the $1\sigma $
statistical uncertainty.

\begin{figure}[f]      
\centerline{\epsfxsize 3.5 truein \epsfbox{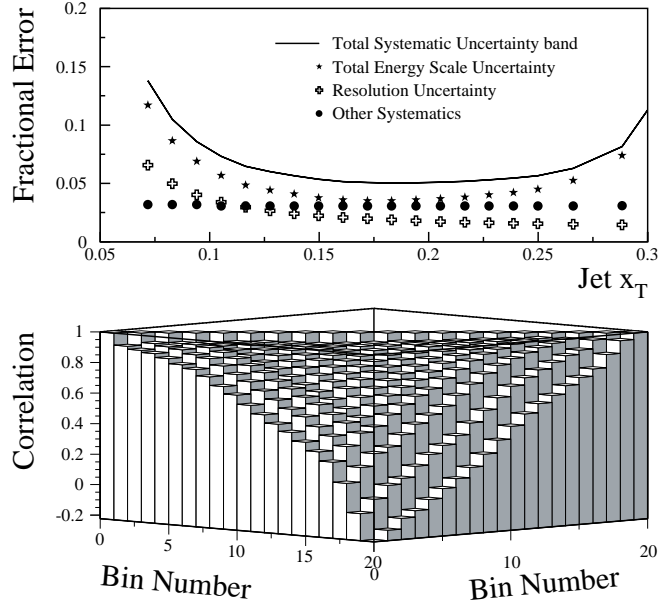}}
\caption[]{(Top) Fractional uncertainty in the ratio of cross sections, displayed
           by component as functions of jet $x_{T}$. (Bottom) Correlations in
           the uncertainty from bin to bin.}
\label{fig:error}
\end{figure}

The primary uncertainty in the ratio of cross sections results from the
energy scale correction, and the remaining uncertainty consists mostly of the
resolution and luminosity uncertainties, as depicted in Fig. \ref{fig:error}.
Although the systematic error in the individual cross sections ranges from
10\% to as much as 30\%, strong numerator--to--denominator correlations in
each bin of the ratio of cross sections reduce the uncertainty to as low as $%
\pm 5\%$.

Because they result primarily from best--fits, the uncertainties associated
with the energy scale and unsmearing corrections do not allow the cross
section ratio to normalize freely. Instead, deviations in one portion of the
distribution increase the probability for deviations in other parts of the 
distribution. The correlation of one bin in the ratio to another determines the
shape evolution of the ratio, given a proposed displacement at a particular
point. The correlation information is represented in the bottom of
Fig. \ref{fig:error}, where the diagonal elements are unity by definition.

\begin{figure}[b]      
\centerline{\epsfxsize 7.0 truein \epsfbox{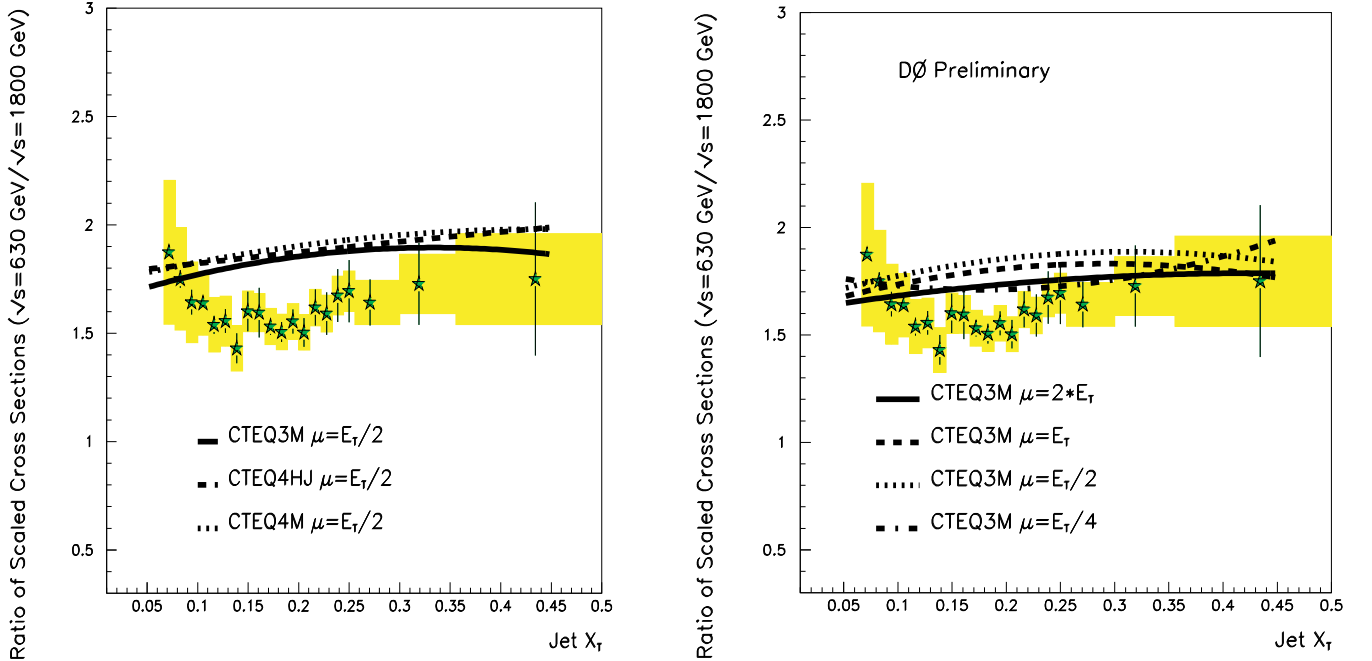}}
\caption[]{The ratio of dimensionless jet cross sections, with NLO QCD
           predictions for: (left) various pdf's and (right) various 
           renormalization scales.}
\label{fig:ratio}
\end{figure}

The observed ratio of scaled cross sections takes a value between 1.4 and
2.0, depending on the position in $x_{T}$. As shown in Fig. \ref{fig:ratio}, 
NLO QCD predictions for the ratio of cross sections lie systematically above 
the data throughout the majority of the measured $x_{T}$ range, most notably in
the range near 0.17, where the ratio has the most statistical power. Many
theoretical choices, including choice of parton distribution function,
minutely change the ratio result; only the renormalization/factorization
scale changes the prediction significantly. The right-hand side of 
Fig. \ref{fig:ratio} compares the
data to predictions with various renormalization scales, 
which were generated with \textsc{Jetrad} 
\cite{jetrad}. 

\begin{figure}[t]      
\centerline{\epsfxsize 3.5 truein \epsfbox{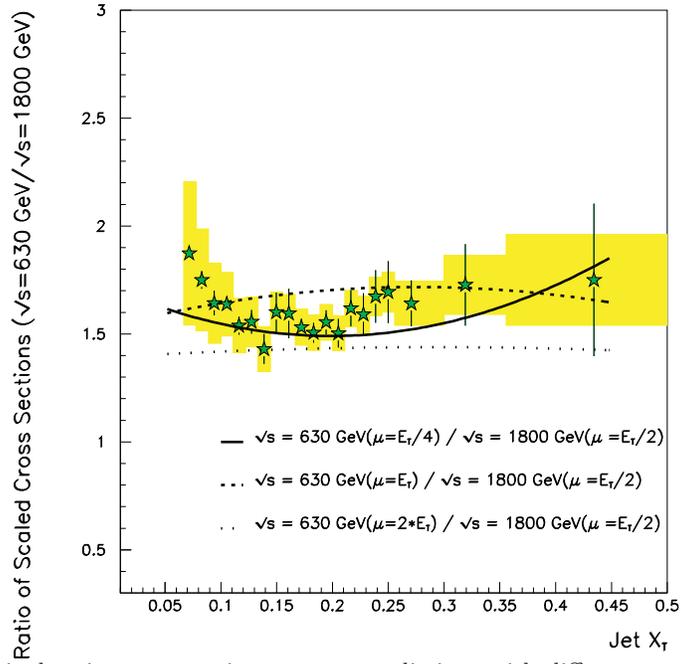}}
\caption[]{The ratio of dimensionless jet cross sections, versus
           predictions with different $\mu$ scales in the numerator
           and the denominator.}
\label{fig:ratio2}
\end{figure}

\begin{table}[t]
\begin{center}
\begin{tabular}{|ccccc|}
\bf{PDF} & $\mu \;$\bf{Scale 630 GeV} & $\mu \;$\bf{Scale 1800
GeV} & $\chi ^{2}$ & prob. \\ \hline
& 2$\cdot $E$_{\mathrm{T}}$ & 2$\cdot $E$_{\mathrm{T}}$ & 29.9 & 7.19\% \\ 
& E$_{\mathrm{T}}$ & E$_{\mathrm{T}}$ & 41.8 & 2.91\% \\ 
CTEQ3M & $\frac{3}{4}$E$_{\mathrm{T}}$ & $\frac{3}{4}$E$_{\mathrm{T}}$ & 51.1
& 0.02\% \\ 
& E$_{\mathrm{T}}/2$ & E$_{\mathrm{T}}/2$ & 50.8 & 0.02\% \\ 
& E$_{\mathrm{T}}/4$ & E$_{\mathrm{T}}/4$ & 30.9 & 5.66\% \\ \hline
CTEQ4HJ & E$_{\mathrm{T}}/2$ & E$_{\mathrm{T}}/2$ & 51.7 & 0.01\% \\ \hline
MRSA$^{\prime }$ & E$_{\mathrm{T}}/2$ & E$_{\mathrm{T}}/2$ & 56.6 & 0.002\%
\\ \hline
CTEQ3M, EKS & E$_{\mathrm{T}}/4$ & E$_{\mathrm{T}}/4$ & 27.9 & 11.0\% \\ 
\hline\hline
& 2$\cdot $E$_{\mathrm{T}}$ & E$_{\mathrm{T}}/2$ & 10.9 & 94.7\% \\ 
CTEQ3M & E$_{\mathrm{T}}$ & E$_{\mathrm{T}}/2$ & 28.5 & 9.84\% \\ 
& E$_{\mathrm{T}}/4$ & E$_{\mathrm{T}}/2$ & 12.1 & 91.3\% \\ 
\end{tabular}
\label{table:results}
\caption{D\O's preliminary $\chi^{2}$ comparisons for the ratio of jet
         cross sections (20 degrees of freedom).}
\end{center}
\end{table}
The matrix of errors and correlations allows a $\chi^{2}$ comparison
between data and theory that can quantitatively describe the probability
that the two curves describe the same parent distribution. The number of data
points, 20, determines the number of degrees of freedom. The results, in
Table~\ref{table:results}, indicate poor agreement between the data and NLO 
QCD for predictions with the same renormalization scale ($\mu$ scale) 
in the numerator and the denominator.  For predictions with different
$\mu$ scales for two $\sqrt{s}$, agreement improves to as much as a 95\% 
probability that the data and prediction represent the same distribution
(Fig. \ref{fig:ratio2}).
The need for different $\mu$ scales may indicate a residual dependence 
of the prediction to higher--order terms in the calculation.

\section{The Preliminary CDF Result}

\begin{figure}[f]      
\centerline{\epsfxsize 3.5 truein \epsfbox{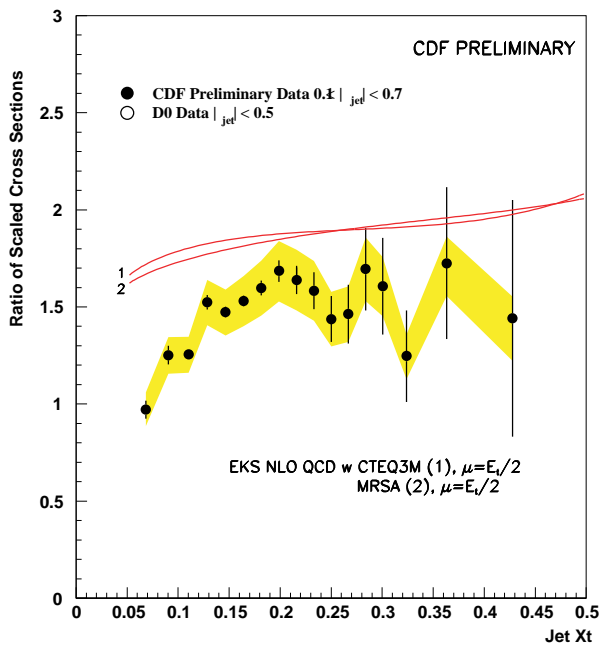}}
\caption[]{CDF's preliminary ratio of dimensionless jet cross sections.}
\label{fig:cdf}
\end{figure}

Shortly after DPF '99, the CDF collaboration updated its preliminary result 
with two major improvements.  First, the ratio of cross sections reflects an
improvement in the available time--integrated luminosity of the 1800 
GeV cross section, which increased from 4,430 nb$^{-1}$ to 87,000 nb$^{-1}$.  
(At 630 GeV, the updated CDF luminosity totals 576~nb$^{-1}$.)
Second, the result \cite{alex} now includes a systematic error estimate
(Figure \ref{fig:cdf}). 

A comparison of Figure \ref{fig:ratio} and Figure \ref{fig:cdf} 
reveals excellent agreement between the experiments above jet $x_T$ of 0.12, 
but sharp disagreement below that value.  Currently, the source of this 
disagreement between experiments is not understood.  The CDF data in 
Figure \ref{fig:cdf} are entirely consistent with the CDF result of 1993 
in Reference \cite{cdf_ratio}.

\section{Conclusions}
The ratio of inclusive jet cross sections provides a strong test of NLO QCD
matrix elements while suppressing the uncertainties that result from pdf's.
These D\O\ and CDF results are drawn from larger data sets than previously
available; thus, they span a larger range of jet $x_T$ and exhibit smaller
uncertainties than any prior measurement.  No quantitative comparison is 
needed to see that the present CDF result exhibits a clear disagreement 
with D\O\ at low values of jet $x_T$.

The $\chi^{2}$ tests of D\O\ indicate that NLO QCD does not model the 
D\O\ data unless a different renormalization scale is introduced for the two 
center--of--mass energies.  Because the renormalization scale has no
physical meaning, the broad range of $\chi^{2}$ probabilities suggest that 
the precision of the data exceeds the precision of the NLO predictions.

\end{document}